\documentclass[12pt]{article}
\pdfoutput=1

\usepackage{color}
\usepackage{amsmath,amssymb,bm,bbm,multirow,float,array,bbold,subfigure}
\usepackage{epsf}
\usepackage{epsfig}
\usepackage{afterpage}
\usepackage{longtable}
\usepackage{booktabs}
\usepackage{caption}
\usepackage[dvipsnames]{xcolor}
\usepackage[linktoc=page,bookmarks=false,colorlinks=false,
linkbordercolor=RoyalBlue,citebordercolor=ForestGreen,urlbordercolor=CornflowerBlue]{hyperref}
\usepackage{latexsym,mathrsfs,dsfont}
\usepackage[normalem]{ulem} 
\usepackage[compress]{cite}
\usepackage{graphicx}
\usepackage{url}
\usepackage{paralist}
\usepackage{siunitx,dcolumn}
\usepackage{slashed}

\renewcommand{\arraystretch}{1.25}

\usepackage{fancyhdr}
\advance \headheight by 15.0truept       

\def \refeq#1{(\ref{#1})}
\def \refsec#1{Section~\ref{#1}}

\def \refapp#1{Appendix~\ref{#1}}
\def \reffig#1{Figure~\ref{#1}}
\def \reftab#1{Table~\ref{#1}}

\newcommand{\be}{\begin{equation}}
\newcommand{\ee}{\end{equation}}

\definecolor{darkgreen}{rgb}{0.0,0.6,0.0}

\DeclareMathOperator{\re}{Re}
\DeclareMathOperator{\im}{Im}

\newcommand{\ord}{\mathcal{O}}

\newcommand{\bra}[1]{\ensuremath{\langle #1 |}}
\newcommand{\ket}[1]{\ensuremath{| #1 \rangle }}

\newcommand{\GeV}{\,\text{GeV}}
\newcommand{\geV}{\text{GeV}}

\newcommand{\tev}{\, \text{TeV}}
\newcommand{\gev}{\, \text{GeV}}

\newcommand{\bsi}{B_6^{(1/2)}}
\newcommand{\bei}{B_8^{(3/2)}}

\def\epe{\varepsilon'/\varepsilon}

\newcommand{\epsK}{\varepsilon_K}

\newcommand{\OmHatEff}{\hat\Omega_\text{eff}}

\newcommand{\alS}{\alpha_s}
\newcommand{\alE}{\alpha_\text{em}}

\newcommand{\muEW}{\mu_\text{EW}}

\newcommand \oL[1]{{\overline{#1}}}
\newcommand \wT[1]{\widetilde{#1}}

\newcommand{\msbar}{\ensuremath{\overline{\text{MS}}}}

\begin{document}

\vspace{-14mm}
\begin{flushright}
  AJB-20-1 \\
  TUM-HEP-1261/20
\end{flushright}

\vspace{6mm}

\begin{center}
{\Large\bf
\boldmath{$\epe$ in the Standard Model at the Dawn of the 2020s}}
\\[12mm]
{\bf \large
  Jason Aebischer${}^a$,
  Christoph Bobeth${}^b$ and
  Andrzej~J.~Buras${}^c$
}
\\[0.8cm]
{\small
${}^a$
  Department of Physics, University of California at San Diego,
  La Jolla, CA 92093, USA
\\[1mm]
  jaebischer@physics.ucsd.edu
\\[2mm]
${}^b$
  Physik Department, TU M\"unchen,
  James-Franck-Stra{\ss}e, D-85748 Garching, Germany
\\[1mm]
  christoph.bobeth@ph.tum.de
\\[2mm]
${}^c$
  TUM Institute for Advanced Study,
  Lichtenbergstr.~2a, D-85748 Garching, Germany
\\[1mm]
  aburas@ph.tum.de
}
\end{center}


\vspace{6mm}

\begin{abstract}
\noindent
We reanalyse the ratio $\epe$ in the Standard Model (SM) using most recent
hadronic matrix elements from the RBC-UKQCD collaboration in combination
with most important NNLO QCD corrections to electroweak penguin contributions
and the isospin-breaking corrections. We illustrate the
importance of the latter by using their latest estimate from chiral
perturbation theory (ChPT) based on the {\em octet} approximation for
lowest-lying mesons and a very recent  estimate in the {\em nonet} scheme
that takes into account the contribution of $\eta_0$. We find
$(\epe)^{(8)}_\text{SM} = (17.4 \pm 6.1) \times 10^{-4}$ and
$(\epe)^{(9)}_\text{SM} = (13.9 \pm 5.2) \times 10^{-4}$, respectively.
Despite a very good agreement with the measured value $(\epe)_\text{exp}
= (16.6 \pm 2.3) \times 10^{-4}$, the large error in $(\epe)_\text{SM}$
still leaves room for significant new physics (BSM) contributions to this
ratio. We update the 2018 master formula for $(\epe)_\text{BSM}$ valid in
any extension beyond the SM without additional light degrees of freedom.
We provide new values of the penguin parameters $\bsi(\mu)$ and $\bei(\mu)$
at the $\mu$-scales used by the RBC-UKQCD collaboration and at lower scales
$\ord(1\gev)$ used by ChPT and DQCD. We present semi-analytic formulae for
$(\epe)_\text{SM}$ in terms of these parameters and $\OmHatEff$ that
summarizes isospin-breaking corrections to this ratio. We stress the
importance of lattice calculations of the $\ord(\alE)$ contributions
to the hadronic matrix elements necessary for the  removal of renormalization
scheme dependence at $\ord(\alE)$ in the present analyses of $\epe$.
\end{abstract}

\setcounter{page}{0}
\thispagestyle{empty}
\newpage

\tableofcontents
\newpage

%
%
%
\section{Introduction}
\label{sec:intro}

The direct CP-violation in $K\to\pi\pi$ decays, represented by the ratio $\epe$,
plays a very important role in the tests of the Standard Model (SM) and more
recently in constraining its possible extensions \cite{Buras:2018wmb}. In the
SM $\epe$ is governed by QCD penguins (QCDP) but receives also an important
contribution from the electroweak penguins (EWP), pointed out already in 1989
\cite{Flynn:1989iu, Buchalla:1989we}, that entering $\epe$ with the opposite
sign to QCDP suppress this ratio significantly. The partial cancellation of these
two contributions in addition to the evaluation of the hadronic matrix elements
of QCDP and EWP operators is the reason why even today a precise prediction for
$\epe$ in the SM is  not available. Yet, significant progress has been made during
the last years and the goal of our paper is to update the SM value of $\epe$
taking into account all available informations both from lattice QCD (LQCD) and
analytic approaches most relevant for the evaluation of the Wilson coefficients
but presently also for the estimate of the isospin-breaking corrections
to the isospin amplitudes.

The situation of $\epe$ in the SM before April 20, 2020 has been summarized
by us in \cite{Aebischer:2019mtr}. In short there are presently three approaches
to calculate hadronic matrix elements entering $\epe$:
\begin{itemize}
\item {\bf Lattice QCD}, lead by the RBC-UKQCD LQCD collaboration. Using their
  results from 2015 for $K\to \pi\pi$ matrix elements \cite{Bai:2015nea,
  Blum:2015ywa} and including isospin-breaking corrections
  from~\cite{Cirigliano:2003nn, Cirigliano:2003gt} as done in \cite{Buras:2015yba,
  Kitahara:2016nld}, leads to a value for $\epe$ in the ballpark of $(1-2)
  \times 10^{-4}$. Although exhibiting a large error of $5 \times 10^{-4}$
  the result lies one order of magnitude below the data. Taking these analyses
  at face value one could talk about an $\epe$ anomaly of at most~$3\,\sigma$.
\item {\bf The Dual QCD (DQCD) approach} \cite{Buras:2014maa, Buras:2015xba, Buras:2016fys},
  which gave a support to these values and moreover provided an \textit{upper
  bound} on $\epe$ in the ballpark of $8\times 10^{-4}$. The main QCD dynamics
  suppressing $\epe$ in this approach is represented by the meson evolution,
  which is necessary to match long-distance contributions to short-distance
  ones. On the other hand it has been argued in \cite{Buras:2016fys} that
  final state interactions (FSI) should have only a minor impact on $\epe$ and
  the quoted bound does not include them.
 \item {\bf Chiral Perturbation theory (ChPT)} \cite{Gisbert:2017vvj,
  Cirigliano:2019cpi, Cirigliano:2019obu} where, using ideas from ChPT, the
  authors found $\epe = (14 \pm 5) \times 10^{-4}$ attributing an
  important role to FSI in this result. While in agreement with the measurement,
  the large uncertainty, that expresses the difficulties in matching
  long-distance and short-distance contributions in this framework, does not
  allow for any clear-cut conclusions.\footnote{See also \cite{Buras:2016fys,
  Buras:2018ozh} for a critical analysis of this approach as used in the context
  of $\epe$.}
\end{itemize}

In view of the fact that LQCD calculations contain both the meson
evolution\footnote{This has been demonstrated for the case of the BSM operators
contributing to $K^0-\bar K^0$ mixing in \cite{Buras:2018lgu}.} and FSI,
while the estimate of $\epe$ in the other two approaches does
not include one of them,
we have recently proposed the optimal strategy for the evaluation of $\epe$
as of 2020 \cite{Aebischer:2019mtr, Buras:2019vik}
\begin{enumerate}
\item Use LQCD  results for hadronic matrix elements of the dominant QCDP and
  EWP operators $Q_6$ and $Q_8$, respectively. They are represented by the
  parameters $\bsi$ and $\bei$ defined in \refsec{sec:2}. On the other hand
  the hadronic matrix elements of $(V-A)\otimes(V-A)$ operators can be
  determined from the experimental data on the real parts of the $K\to\pi\pi$
  amplitudes as performed in \cite{Buras:1993dy, Buras:2015yba}. In fact this
  procedure has been recently adopted with slight modifications by the RBC-UKQCD
  collaboration \cite{Abbott:2020hxn} with the goal to decrease their errors.
  This procedure is clearly legitimate when testing the consistency of the
  SM with the data, if the reduction of the uncertainties is significant.
  But the RBC-UKQCD analysis shows only a little gain and therefore we will
  not use it in the present work, but rather base the full analysis on {\em all}
  hadronic matrix elements from RBC-UKQCD that are absolutely free from new
  physics. We will only use the experimental values of $\re A_{0,2}$ in the
  basic formula for $\epe$ because they automatically take possible NP
  contributions into account.
\item Include isospin-breaking corrections from ChPT \cite{Cirigliano:2019cpi}
  that are compatible due to large uncertainties with the results obtained
  already 33 years ago in \cite{Buras:1987wc}. Very recently the latter analysis
  has been updated \cite{Buras:2020pjp} and we will include these new
  findings as well.
\item Include NNLO QCD contributions to EWP in \cite{Buras:1999st} thereby
  reducing the unphysical scale and renormalization scheme dependences
  in the matching at $\mu_W = \ord(m_W)$, with the largest part
  due to the top-quark mass. The removal of the dependence on $\mu_c$ at
  NNLO has still to be done, see also the next point.
\item Take into account NNLO QCD contributions to QCDP \cite{Cerda-Sevilla:2016yzo,
  Cerda-Sevilla:2018hjk}. This reduces the left-over renormalization scale
  uncertainties present at the NLO level, in particular those due
  to the matching  scale $\mu_c$.
\end{enumerate}

Recently significant progress in the estimate of $\epe$ in the SM has been
made through the improved values of the $K\to\pi\pi$ hadronic matrix elements
presented by the RBC-UKQCD collaboration \cite{Abbott:2020hxn}. Not only
statistical errors have been significantly decreased but also a better
agreement with the experimental values of the $\pi\pi$ strong interaction phases
$\delta_{0,2}$ has been obtained. The RBC-UKQCD collaboration, using their
new results for the hadronic matrix elements and known Wilson coefficients
at the NLO level \cite{Buras:1991jm, Buras:1992tc, Buras:1992zv, Buras:1993dy,
Ciuchini:1992tj, Ciuchini:1993vr} but not accounting for isospin-breaking
corrections, finds \cite{Abbott:2020hxn}
\begin{align}
  \label{RBCUKQCD}
  (\epe)_\text{SM} &
  = (21.7 \pm 8.4) \times 10^{-4} \,,\qquad (\text{RBC-UKQCD}-2020)
\end{align}
to be compared with the experimental world average from NA48
\cite{Batley:2002gn} and KTeV \cite{AlaviHarati:2002ye, Worcester:2009qt}
collaborations,
\begin{align}
  \label{EXP}
  (\epe)_\text{exp} &
  = (16.6 \pm 2.3) \times 10^{-4} \,.
\end{align}

While the result in \eqref{RBCUKQCD} is in full agreement with the experimental
value in  \eqref{EXP} the theoretical error of $39\%$ does not allow for clear
cut conclusions whether some amount of new physics contributions is present in
$\epe$ or not. The same is the case of the earlier updated ChPT analysis
\cite{Cirigliano:2019cpi}, which resulted in
\begin{align}
  \label{Pich}
  (\epe)_\text{SM} &
  = (14 \pm 5) \times 10^{-4} \,,\qquad (\text{ChPT}-2019),
\end{align}
with an error of $36\%$, very close to the LQCD one. But it should
be remarked that with the present best values of the CKM parameters as used
by us the central value in \eqref{Pich} would be raised to
$15.0 \times 10^{-4}$.

Despite large errors both results deviate significantly from the DQCD
values of $\epe$ in the ballpark of $5\times 10^{-4}$ stressed in particular
in \cite{Buras:2018ozh}. While there is no question about that meson evolution
necessary for a proper matching between Wilson coefficients and hadronic
matrix elements at scales $\ord(1\gev)$ must play a role in the evaluation
of $\epe$ it appears from present RBC-UKQCD results that precisely in the
case of the matrix element of the $Q_6$ operator its suppression is
overcompensated by other QCD dynamics which was hidden due to the
contamination of the excited $\pi\pi$ states present in their 2015 analysis.
It has been removed in the latest analysis. In fact as we will see soon
the value of $\epe$ obtained using the optimal procedure with hadronic
matrix elements from \cite{Abbott:2020hxn}, agrees very well with the one
advocated in \cite{Cirigliano:2019cpi} and given in \eqref{Pich}. Yet, it
is not evident at present that FSI, as claimed by ChPT experts, are responsible
for this agreement. Possibly other dynamical QCD effects apparently not taken
into account both in the ChPT and DQCD approaches are responsible for the
enhancement of $\epe$ relative to DQCD expectations.\footnote{See also
the discussion in \cite{Buras:2020pjp} on this point.} However, a clear-cut
conclusion on this issue is difficult because of rather different techniques
that are used in these three approaches. The fact that the central value
 in \eqref{Pich} differs significantly from the central LQCD value in
\eqref{RBCUKQCD} is dominantly due to the omission of isospin-breaking
effects in the RBC-UKQCD prediction that are included in \eqref{Pich}.

Even if the new improved calculation of $K\to\pi\pi$ hadronic matrix elements
in \cite{Abbott:2020hxn} is an important advance towards the accurate
calculation of $\epe$, the result in \refeq{RBCUKQCD} does not represent
the present SM value of $\epe$ properly. Indeed, as we emphasized in
\cite{Aebischer:2019mtr} the hadronic matrix elements in question are only
a part of the $\epe$ story. The three additional advances, listed in the
context of the optimal strategy, that are not taken into account in the
result in (\ref{RBCUKQCD}) are also important, in particular because they
all lower the value of $\epe$. As we will demonstrate below, the final
result for $\epe$ differs significantly from the one obtained by the
RBC-UKQCD collaboration. Indeed after including isospin-breaking
effects from \cite{Cirigliano:2019cpi} that include the effects
from the {\em octet} of lowest-lying mesons and NNLO QCD corrections to EWP
contributions, we find using the hadronic matrix elements of RBC-UKQCD
\begin{align}
  \label{ABBG}
  (\epe)^{(8)}_\text{SM}
  = (17.4 \pm 6.1) \times 10^{-4} \,.
\end{align}
On the other hand including the singlet $\eta_0$ in this estimate one arrives
at \cite{Buras:2020pjp}
\begin{align}
  \label{BG20}
  \boxed{
  (\epe)^{(9)}_\text{SM}
  = (13.9 \pm 5.2) \times 10^{-4}} \,.
\end{align}
Both results agree very well with experiment and with the ChPT expectations
but in view of our comments on the ChPT analysis are on a more solid footing.
We expect further reduction of $\epe$  by roughly $(5-10)\%$ when NNLO QCD
corrections to QCD penguin contributions will be taken into account
\cite{Cerda-Sevilla:2016yzo, Cerda-Sevilla:2018hjk}. We look forward
to the final results of these authors.

Our paper is organized as follows. In \refsec{sec:2}, after recalling a number
of basic formulae, we determine the parameters $\bsi$ and $\bei$ using the
recent RBC-UKQCD results and compare them with the expectations from
ChPT \cite{Cirigliano:2019cpi} and DQCD \cite{Buras:2015xba, Buras:2016fys}.
It turns out that while there is a good agreement on the value of $\bsi$
between LQCD and ChPT, the rather precise value of $\bei$ from RBC-UKQCD is
by a factor of $1.5$ larger than the ChPT one when both are evaluated at
$\mu = 1\GeV$. On the contrary, while there is a good agreement on the value
of $\bei$ between LQCD and DQCD \cite{Buras:2015xba}, the most recent value
of $\bsi$ from RBC-UKQCD is by a factor of two larger than the values quoted
in \cite{Buras:2015xba, Buras:2016fys}. We close this section with an updated
formula for $\epe$ in terms of $\bsi$ and $\bei$. In \refsec{sec:3} we derive
the results in \eqref{ABBG} and \eqref{BG20} which take into account the
updated isospin-breaking effects \cite{Cirigliano:2019cpi, Buras:2020pjp}
and also NNLO QCD corrections to EWP contributions \cite{Buras:1999st}. We
also perform a  detailed anatomy of various contributions. In \refsec{sec:NP}
we update the BSM master formula for $\epe$ \cite{Aebischer:2018quc,
Aebischer:2018csl} in view of the new RBC-UKQCD results. A brief summary and
an outlook are given in \refsec{sec:4}. Some additional information on the
numerical analysis are given in appendices. This includes the values of the
hadronic matrix elements from RBC-UKQCD and  the Wilson coefficients at
various scales. We discuss in detail the effect of isospin-breaking
corrections present in the renormalization group (RG) flow on $\epe$ in
\refapp{app:RG-isopin-breaking}.

%
%
%
\section{Basic formulae}
\label{sec:2}

%
\subsection{Preliminaries}

The amplitudes for $K^0\to (\pi\pi)_I$, with $I=0,2$ denoting strong isospin
of the final state, are given as
\begin{align}
  \label{eq:def-A0}
  A_0 &
  = \mathcal{N}_{\Delta S = 1} \sum_{i=1}^{10}
    \big[z_i(\mu) + \tau y_i(\mu) \big] \langle Q_i(\mu)\rangle_0 \,,
\\
  \label{eq:def-A2}
  A_2 &
  = \mathcal{N}_{\Delta S = 1} \sum_{i=1}^{10}
    \big[ z_i(\mu) + \tau y_i(\mu)\big] \langle Q_i(\mu)\rangle_2 \,,
\end{align}
where $z_i(\mu)$ and $y_i(\mu)$ are the $\Delta S = 1$ Wilson coefficients
and $\langle Q_i(\mu)\rangle_{0,2}$ the hadronic matrix elements of the
operators $Q_i$, both in the \msbar{} scheme at the low-energy factorization
scale $\mu$ in the $N_f = 3$ flavour theory \cite{Buras:1993dy}. By convention
the strong phase shifts $\delta_{0,2}$ are not included in $A_{0,2}$, and
therefore the $\langle Q_i(\mu)\rangle_{0,2}$ are real-valued. Further
\begin{align}
  \mathcal{N}_{\Delta S = 1} &
  = \frac{G_F}{\sqrt{2}} V_{us}^* V_{ud}^{} ,
&
  \tau &
  = - \frac{V_{ts}^*\, V_{td}^{}}{V_{us}^* V_{ud}^{}} .
\end{align}
The real parts $\re A_{0,2}$ are given entirely by the $z_i$, because the
$y_i$ are strongly suppressed by $\tau \sim \mathcal{O}(10^{-3})$, on the
other hand the imaginary parts $\im A_{0,2} \propto \im (V_{ts}^*\, V_{td}^{})$
and depend only on~$y_i$. The Wilson coefficients of the QCD penguin
(QCDP) operators $i=3,\ldots,6$ are usually larger compared to those
of the electroweak penguin (EWP) operators $i=7,\ldots,10$, as can
be seen in \reftab{tab:wc-values}.

The scheme and scale dependences cancel between the Wilson coefficients
and the matrix elements individually in $A_0$ and $A_2$. We will take
advantage of this freedom to use different scales $\mu_0$ and $\mu_2$ in
the evaluation of $A_0$ and $A_2$, respectively. In particular we choose
the values at which the RBC-UKQCD lattice collaboration presents their
results of the $I=0$~\cite{Abbott:2020hxn} and $I=2$ \cite{Blum:2015ywa}
matrix elements. There are only seven linearly independent
$\langle Q_i(\mu) \rangle_0$ and three linearly independent
$\langle Q_i(\mu)\rangle_2$ in the $N_f = 3$ flavour theory
\cite{Buras:1993dy, Buras:2015yba}, since RBC-UKQCD work in
the isospin-symmetric limit, where also QED corrections are not included
yet.

We remind that the amplitudes $A_{0,2}$ and the strong phase shifts
$\delta_{0,2}$ are related to the decay amplitudes relevant for $\epe$
as follows
\begin{equation}
\begin{aligned}
  A(K^0 \to \pi^+\pi^-) &
  = \frac{1}{h} \Big[ A_0 e^{i \delta_0}
                     + \frac{1}{\sqrt{2}} A_2 e^{i \delta_2} \Big] ,
& \;\;\;
  A(K^0 \to \pi^0\pi^0) &
  = \frac{1}{h} \Big[ A_0 e^{i \delta_0}
                     - \sqrt{2}\, A_2 e^{i \delta_2} \Big] , &
\end{aligned}
\end{equation}
with the experimental values of $A_{0,2}$ for $h = 1$ given in
\reftab{tab:num-input}, whereas RBC-UKQCD works with the convention
$h = \sqrt{3/2}$. These relations are valid also in the
presence of finite QED corrections, as long as virtual infrared-divergent
contributions, and also Coulomb corrections, are properly subtracted and
combined with real photon radiation \cite{Cirigliano:2003gt} when
determining the amplitudes and phases from data.

%
\subsection[Basic formula for $\epe$]
{Basic formula for \boldmath{$\epe$}}

As in \cite{Buras:2015yba}, our starting expression is the formula
\begin{align}
  \label{eq:eprime}
  \frac{\varepsilon'}{\varepsilon} &
  = -\,\frac{\omega_+}{\sqrt{2}\,|\varepsilon_K|}
  \left[\, \frac{\im \wT{A}_0}{\re A_0}\, (1 - \OmHatEff)
       -   \frac{1}{a} \, \frac{\im A_2}{\re A_2} \,\right],
\end{align}
where \cite{Cirigliano:2003gt, Cirigliano:2019cpi}
\begin{align}
  \label{OM+}
  \omega_+ &
  = a\,\frac{\re A_2}{\re A_0}
  = (4.53 \pm 0.02) \times 10^{-2} ,
&
  a & = 1.017 .
\end{align}
Here $a$ and $\OmHatEff$ summarise isospin-breaking corrections. The latter
include strong isospin violation $(m_u \neq m_d)$, the correction to the isospin
limit coming from $\Delta I=5/2$ transitions and electromagnetic corrections
as first summarized in \cite{Cirigliano:2003nn, Cirigliano:2003gt} and
recently updated in \cite{Cirigliano:2019cpi}
\begin{align}
  \label{eq:OmEff-8}
  \OmHatEff & = \OmHatEff^{(8)} = (17.0 \pm 9.1) \times 10^{-2} .
\end{align}
These analyses are based on the so-called {\em octet scheme} which
includes only the octet of the lowest-lying pseudoscalar mesons. The inclusion
of the singlet $\eta_0$ in the {\em nonet scheme} has been known already for
33 years \cite{Donoghue:1986nm, Buras:1987wc} to give stronger suppression of
$\epe$ through the $\eta-\eta^\prime$ mixing, but only very recently this
estimate has been updated and put on a more solid basis than it was possible
in 1987. With
\begin{align}
  \label{eq:OmEff-9}
  \OmHatEff & = \OmHatEff^{(9)} = (29 \pm 7) \times 10^{-2} ,
\end{align}
the role of isospin-breaking effects is enhanced relative to the ChPT
estimate in \eqref{eq:OmEff-8}.

The inclusion of
the isospin-breaking corrections requires a modification in the evaluation
of the $\im A_0$ part in $\epe$ as follows \cite{Buras:2015yba}
\begin{align}
  \label{eq:def-im-tildeA0}
  \im A_0 \quad\to\quad
  \im \wT{A}_0 &
  = \mathcal{N}_{\Delta S = 1} \im \tau \left[
      \sum_{i=3}^{6}  y_i(\mu) \langle Q_i(\mu)\rangle_0
    + \sum_{i=7}^{10} \frac{y_i(\mu) \langle Q_i(\mu)\rangle_0}
                           {a (1 - \OmHatEff)}\right] ,
\end{align}
such that only leading isospin-breaking corrections are included.

A strong reduction of the uncertainty of $\epe$ can be achieved firstly
\cite{Buras:1993dy} by the use of the experimental values of $\re A_{0,2}$
in the denominators of \eqref{eq:eprime}. Secondly, the real parts of the
relations \eqref{eq:def-A0} and \eqref{eq:def-A2} allow to eliminate
one $\langle Q_j(\mu_0)\rangle_0$ and one $\langle Q_k(\mu_2)\rangle_2$,
respectively, in favour of the measured values of $\re A_0$ and $\re A_2$,
respectively. These can then be used in the numerators $\im \wT{A}_0$ and
$\im A_2$, as proposed in \cite{Buras:2015yba}. The particular choice of $j$
and $k$ is subject to optimisation. However, as already announced
previously we will not use this procedure here.

The real parts of the isospin amplitudes $A_{0,2}$ in \eqref{eq:eprime}
are then extracted from the branching ratios on $K\to\pi\pi$ decays in
the isospin limit. In the limit $a = 1$ and $\OmHatEff = 0$ the formula in
\eqref{eq:eprime} reduces to the one used by RBC-UKQCD \cite{Abbott:2020hxn},
where all isospin breaking-corrections except for EWP contributions at
the NLO level have been set to zero.

%
\subsection[Extracting $\bsi$ and $\bei$ from LQCD]
{Extracting \boldmath{$\bsi$} and \boldmath{$\bei$} from LQCD}
\label{sec:bag-factors}

In the past the so-called bag factors have been frequently used in
phenomenological analyses and it is interesting to provide their values
in view of the updated $I=0$ matrix elements. The $\bsi$ and $\bei$
parameters, that enter the formula \refeq{eq:semi-num-1}, are defined
as follows
\begin{align}
  \label{eq:Q60}
  \langle Q_6(\mu) \rangle_0 &
  = -\,4 h \left[\frac{m_K^2}{m_s(\mu) + m_d(\mu)}\right]^2 (F_K - F_\pi)
    \,\bsi ,
\\
  \label{eq:Q82}
  \langle Q_8(\mu) \rangle_2 &
  = \sqrt{2} h
  \left[ \frac{m_K^2}{m_s(\mu) + m_d(\mu)} \right]^2 F_\pi \,\bei ,
\end{align}
with \cite{Buras:1985yx,Buras:1987wc}
\begin{align}
  \label{LN}
  \bsi & = \bei = 1\,,
\end{align}
in the large-$N$ limit. We have introduced the factor $h$ in order to emphasize
different normalizations of these matrix elements present in the literature.

We find from the latest RBC-UKQCD results for $I=0$ \cite{Abbott:2020hxn}
matrix elements at the scales $\mu= 1\GeV$, $\mu = \mu_c = 1.3\GeV$ and
$\mu = \mu_0 = 4.006\GeV$
\begin{equation}
\begin{aligned}
  \label{eq:Lbsi}
  \bsi(1.0\GeV) &
  = 1.49 \pm 0.11|_\text{stat} \pm 0.23 |_\text{syst}
  = 1.49 \pm 0.25,
\\
  \bsi(\mu_c) &
  = 1.36 \pm 0.10|_\text{stat} \pm 0.21 |_\text{syst}
  = 1.36 \pm 0.23,
\\
  \bsi(\mu_0) &
  = 1.11 \pm 0.08|_\text{stat} \pm 0.18 |_\text{syst}
  = 1.11 \pm 0.20,
\end{aligned}
\end{equation}
and for $I=2$ from \cite{Blum:2015ywa} for $\mu=1\GeV$, $\mu_c = 1.3 \GeV$ and
$\mu_2 = 3.0 \GeV$
\begin{equation}
\begin{aligned}
  \bei(1.0\GeV) &
  = 0.85 \pm 0.02|_\text{stat} \pm 0.05 |_\text{syst}
  = 0.85 \pm 0.05,
\\
  \bei(\mu_c) &
  = 0.79 \pm 0.02|_\text{stat} \pm 0.05 |_\text{syst}
  = 0.79 \pm 0.05,
\\
  \bei(\mu_2) &
  = 0.70 \pm 0.02|_\text{stat} \pm 0.04 |_\text{syst}
  = 0.70 \pm 0.04,
\end{aligned}
\end{equation}
to be compared with the 2015 values $\bsi(\mu_c) = 0.57 \pm 0.19$ and
$\bei(\mu_c) = 0.76 \pm 0.05$ from RBC-UKQCD \cite{Bai:2015nea, Blum:2015ywa}.
In principle only\footnote{Note though that the used input for quark masses
has been updated here, see \reftab{tab:num-input}. The associated uncertainties
are not included, because in the expressions for $\epe$ the dependence on
these parameters cancels.} the central value of $\bsi$ has been changed by a
factor of more than two, but with slightly larger uncertainty, which
would correspond to a $2.6\,\sigma$ discrepancy. However, in
view that the systematic uncertainty of the 2015 results for the $I=0$
matrix elements has been underestimated \cite{Abbott:2020hxn}, the uncertainty
quoted for the 2015 result of $\bsi(\mu_c)$ must not be taken at face value
anymore.

The new value of $\bsi$ is in the ballpark of values advocated in
\cite{Cirigliano:2019cpi}, but it is unclear to us at present whether this
is a numerical coincidence or due to FSI dynamics. Moreover, the
large uncertainty in the value of $\bsi$ does not yet rule out the values
of $\bsi < 1.0$ as expected from the DQCD approach \cite{Buras:2015xba}.
Similar, the decrease of both parameters with increased $\mu$, pointed out
already in \cite{Buras:1993dy} and seen above, is also present below
$1\GeV$ within the DQCD allowing smooth matching between hadronic matrix
elements and Wilson coefficients. On the other hand it turns out that while
there is a good agreement on the value of $\bei$ between LQCD and
DQCD \cite{Buras:2015xba}, its rather precise value from RBC-UKQCD is by
a factor of $1.5$ larger than the ChPT one, in the ballpark of $0.55$,
when both are evaluated at $\mu = 1\GeV$.

\subsection[An analytic formula for $\epe$]
{An analytic formula for \boldmath{$\epe$}}
\label{sec:analytic-formula}

As is well-known and shown also in the full analysis later, $\epe$ is
strongly dominated by the two terms $\propto \langle Q_6 \rangle_0 \sim \bsi$
and $\propto \langle Q_8 \rangle_2 \sim \bei$. For convenience we provide a
semi-analytic result of $\epe$ in terms of these two parameters. Contrary to
\cite{Buras:2015yba, Aebischer:2019mtr}, we evaluate $A_0$ and $A_2$ at
the two different scales $\mu_0$ and $\mu_2$ and use now for the remaining
matrix elements the RBC-UKQCD results. Then
\begin{align}
  \label{eq:semi-num-1}
  \frac{\varepsilon'}{\varepsilon} & =
  \im \lambda_t \cdot \left[
    a (1 - \OmHatEff) \left(a^\text{QCDP} + a_6^{(1/2)} \bsi \right)
    - a^\text{EWP} - a_8^{(3/2)} \bei
  \right]\,,
\end{align}
with the coefficients given in \reftab{tab:semi-analytic} for various
choices of $(\mu_0,\, \mu_2)$. The numerical input of the various
parameters entering \eqref{eq:semi-num-1} is given in \reftab{tab:num-input}
and details on the Wilson coefficients at scales $\mu_{0,2}$ are collected in
\refapp{app:wilson-coeffs}. The quark masses in \eqref{eq:Q60} and \eqref{eq:Q82}
have been calculated as well at the two scales $\mu_0$ and $\mu_2$, respectively.
The coefficients $a_i^j$ are comparable to \cite{Aebischer:2019mtr}, but differ
because of the updated values for the remaining $I=0$ matrix elements and changed
values of the down- and strange-quark masses. Note that $a^\text{EWP}$
contains the $I=0$ and $I=2$ contributions of the EWPs at the scales
$(\mu_0,\, \mu_2) = (4.006,\, 3.0)\GeV$, where the QCDP matrix elements for $I=2$
are zero because the lattice calculation is done in the isospin limit.
In general, when using the RG equations to evolve these matrix elements
to different scales $\mu_{0,2}$, the isospin-breaking in quark charges in the
RG flow lead to nonvanishing $I=2$ QCDP matrix elements that would
also contribute to $a^\text{EWP}$. As explained in more detail in
\refapp{app:RG-isopin-breaking}, we evolve the matrix elements of the operators
from the initial scales $\mu_{0,2}$ to the scales $\mu = 1.3\GeV$ and $1.0\GeV$
only with NLO QCD RG equations instead of NLO QCD$\,\times\,$QED, which
maintains isospin relations for these matrix elements.

\begin{table}
\centering
\renewcommand{\arraystretch}{1.4}
\begin{tabular}{|l|rrr|}
\hline
  $(\mu_0,\,\mu_2)$[GeV] & $(1.0, \, 1.0)$ & $(1.3,\,1.3)$ & $(4.006,\, 3.0)$
\\
\hline\hline
  $\bsi(\mu_0)$ & $1.49 \pm 0.25$ & $1.36 \pm 0.23$ & $1.11 \pm 0.20$
\\
  $\bei(\mu_2)$ & $0.85 \pm 0.05$ & $0.79 \pm 0.05$ & $0.70 \pm 0.04$
\\
\hline
  $m_d\,$[MeV]  &   $6.37$ &   $5.52$ & $(3.88,\, 4.16)$
\\
  $m_s\,$[MeV]  & $125.48$ & $108.81$ & $(76.50,\, 81.89)$
\\
\hline
  $a^\text{QCDP}(\mu_0)$       & $-2.86$ & $-3.37$ & $-5.64$
\\
  $a_6^{(1/2)}(\mu_0)$         & $15.15$ & $16.98$ & $22.77$
\\
  $a^\text{EWP}(\mu_0, \mu_2)$ & $-2.02$ & $-2.12$ & $-2.27$
\\
  $a_8^{(3/2)}(\mu_2)$         &  $8.00$ &  $8.79$ &  $9.85$
\\
\hline
  $10^4 \times (\epe)^{(8)}$     & $17.2$ & $17.1$ & $17.3$
\\
  $10^4 \times (\epe)^{(9)}$     & $13.7$ & $13.7$ & $13.9$
\\
\hline
\end{tabular}
\caption{\small
  Coefficients entering the semi-analytic formula \eqref{eq:semi-num-1}, when the
  amplitudes $A_0$ and $A_2$ are evaluated at the scales $\mu_0$ and $\mu_2$,
  respectively, for different choices of $(\mu_0, \mu_2)$.
  The central values for $(\epe)^{(8,9)}$ with these approximate
  formulas are given in the last two lines.
}
  \label{tab:semi-analytic}
\end{table}

%
%
%
\section{\boldmath{$\epe$} in the Standard Model}
\label{sec:3}

\begin{table}
\centering
\renewcommand{\arraystretch}{1.3}
\resizebox{\columnwidth}{!}{
\begin{tabular}{|llllll|}
\hline
  Parameter
& Value
& Ref.
&  Parameter
& Value
& Ref.
\\
\hline\hline
  $G_F$                   & $1.166379 \times 10^{-5} \GeV^{-2}$  & \cite{Tanabashi:2018oca}
& & &
\\
\hline
  $\lambda$               & $0.22453(44)$          & \cite{Tanabashi:2018oca}
& $A$                     & $0.836(15)$            & \cite{Tanabashi:2018oca}
\\
  $\oL{\rho}$             & $0.122(^{+18}_{-17})$  & \cite{Tanabashi:2018oca}
& $\oL{\eta}$             & $0.355(^{+12}_{-11})$  & \cite{Tanabashi:2018oca}
\\
  $V_{ud}$                & $0.97446(10)$          &
& $V_{td}^{} V_{ts}^*$    & $[-3.40(15) + i\, 1.45(8)]\times 10^{-4}$ &
\\
  $V_{us}$                & $0.22453(45)$            &
& $\tau$                  & $[\,15.58(67) - i\, 6.62(35)]\times 10^{-4}$ &
\\
\hline
  $\re A_0|_\text{exp}$   & $27.04(1) \times 10^{-8}$ GeV  & \cite{Cirigliano:2011ny}
& $\epsK$                 & $0.002228(11)$        & \cite{Tanabashi:2018oca}

\\
  $\re A_2|_\text{exp}$   & $1.210(2) \times 10^{-8}$ GeV  & \cite{Cirigliano:2011ny}
& $m_K$                   & $497.614$ MeV         & \cite{Tanabashi:2018oca}
\\
\hline
  $F_\pi$                 & $130.41(20)$ MeV     & \cite{Tanabashi:2018oca}
& $m_d(2\GeV)$            &  $4.67(9)$ MeV       & \cite{Aoki:2019cca}
\\
  $F_K/F_\pi$             & $1.194(5)$           & \cite{Aoki:2019cca}
& $m_s(2\GeV)$            & $92.0(1.1)$ MeV      & \cite{Aoki:2019cca}
\\
\hline
\end{tabular}
}
\caption{\small
  Numerical input: The CKM elements and combinations thereof and the
  uncertainties are derived from Wolfenstein parameters from PDG 2019.
  The experimental results for $K\to \pi\pi$ amplitudes $\re A_{0,2}|_\text{exp}$
  are for normalization $h=1$. The \msbar{} quark masses are FLAG averages
  for $N_f = 2 + 1$ from \cite{Blum:2014tka, Durr:2010vn, Durr:2010aw,
  McNeile:2010ji, Bazavov:2009fk, Fodor:2016bgu}.
}
  \label{tab:num-input}
\end{table}

The new results for the $I=0$ matrix elements from RBC-UKQCD imply a
modification of $\epe$ in the SM relative to those values presented
in 2015 in \cite{Bai:2015nea, Buras:2015yba, Kitahara:2016nld}, taking
into account additional advances listed in \refsec{sec:intro}.
Here we include the isospin-breaking corrections $\OmHatEff$
and NNLO QCD corrections to EWPs calculated in \cite{Buras:1999st}.
Both contributions lead to a considerable reduction of $\epe$, as discussed
previously \cite{Aebischer:2019mtr}. Note that the RBC-UKQCD collaboration
\cite{Abbott:2020hxn} prefers to use the magnitude of the isospin-breaking
corrections from ChPT in the octet scheme~\eqref{eq:OmEff-8}
exclusively as an estimate of their size, thereby introducing an additional large
uncertainty in $\epe$. In contrast to previous predictions \cite{Buras:2015yba,
Aebischer:2019mtr}, here we use in obtaining the final result for $\epe$
directly the LQCD values of matrix elements $\langle Q_i(\mu_0) \rangle_0$
and $\langle Q_i(\mu_2) \rangle_2$. For the interested readers, we provided
the updated values of the two most important bag factors $\bsi$ and $\bei$
in \refsec{sec:bag-factors}.

\begin{figure}
\centering
  \includegraphics[width=0.46\textwidth]{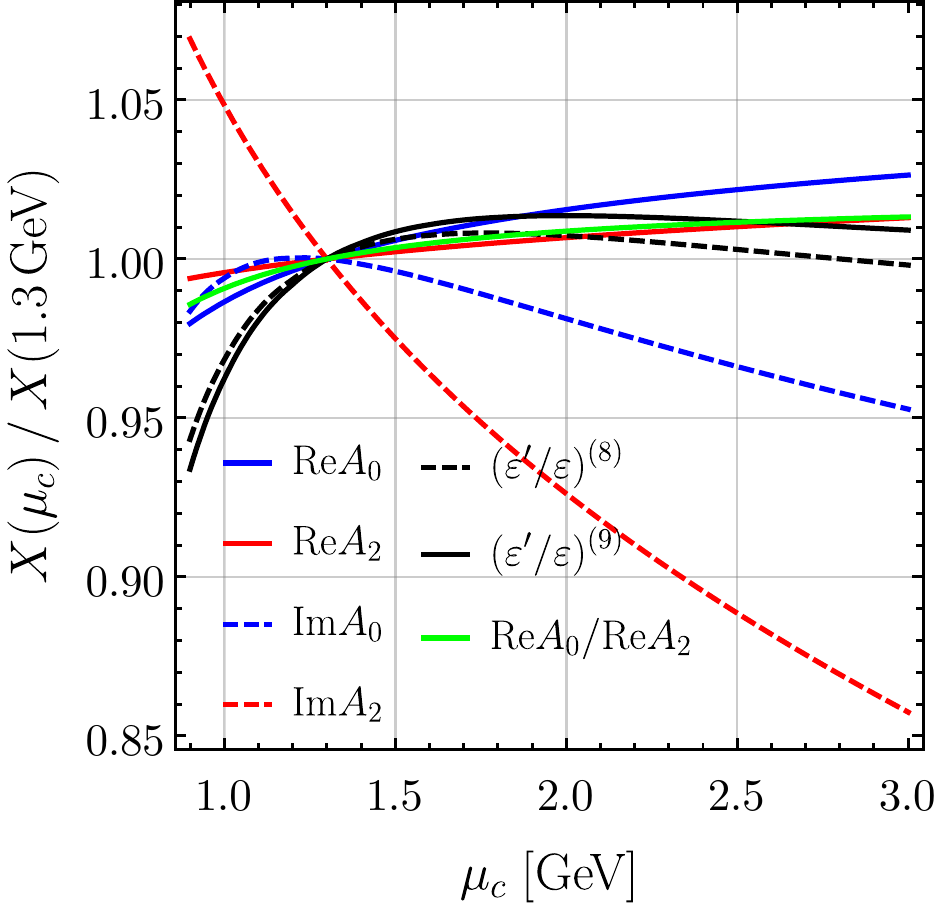}
  \hskip 0.05\textwidth
  \includegraphics[width=0.45\textwidth]{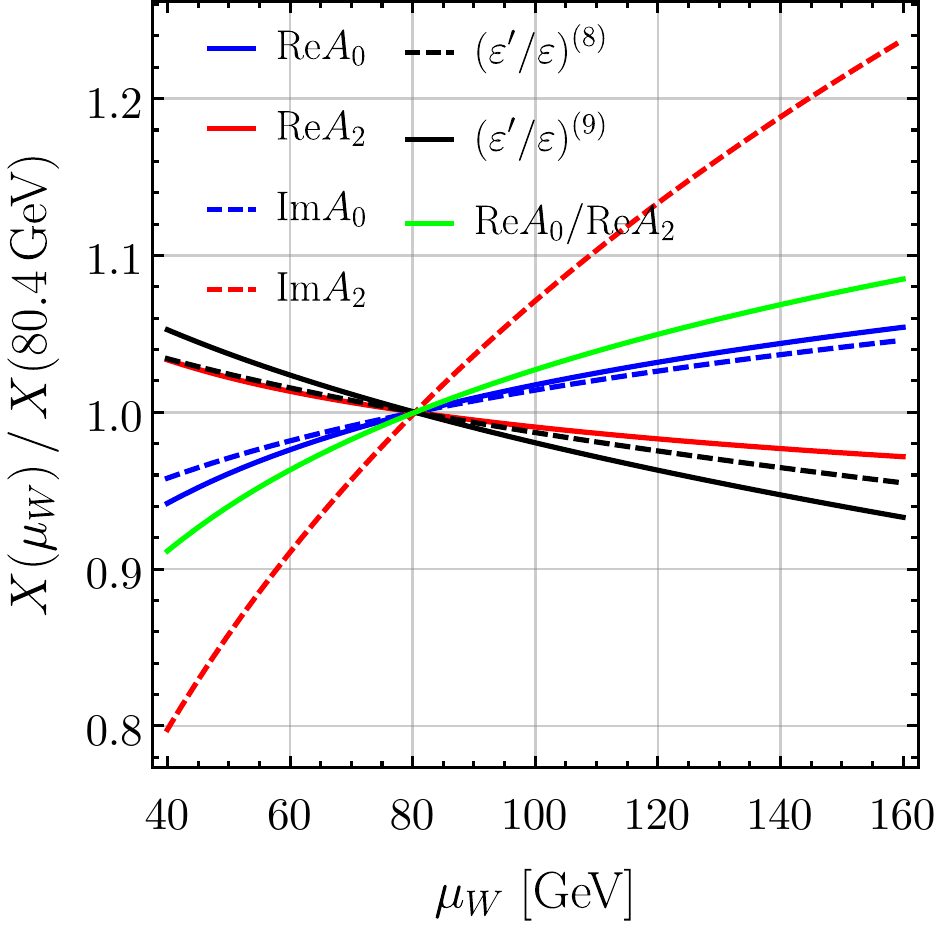}
\caption{\small The $\mu_c$ dependence [left] and $\mu_W$ dependence
  [right] at NLO accuracy of the various quantities
  normalized to their value at $\mu_c = 1.3\GeV$ and $\mu_W = 80.4\GeV$, respectively.}
  \label{fig:mu-dep}
\end{figure}

We find for the amplitudes ($h=1$)
\begin{align}
  \re A_0 &
  = \left(24.63
  \pm 2.65 \big|^\text{ME}_\text{stat}
  \pm 3.87 \big|^\text{ME}_\text{syst}
  \; {}^{+0.63}_{-0.33} \big|_{\mu_c}
  \; {}^{+1.08}_{-0.97} \big|_{\mu_W} \right) \times 10^{-8} \GeV ,
\\
  \re A_2 &
  = \left(\;1.23\,
  \pm 0.03 \big|^\text{ME}_\text{stat}
  \pm 0.07 \big|^\text{ME}_\text{syst}
  \; {}^{+0.02}_{-0.01} \big|_{\mu_c}
  \; {}^{+0.03}_{-0.03} \big|_{\mu_W} \right) \times 10^{-8} \GeV ,
\end{align}
and
\begin{align}
  \im A_0 &
  = \left(-5.74
  \pm 0.53 \big|^\text{ME}_\text{stat}
  \pm 0.90 \big|^\text{ME}_\text{syst}
  \pm 0.30 \big|_\text{CKM}
  \; {}^{+0.00}_{-0.26} \big|_{\mu_c}
  \; {}^{+0.21}_{-0.17} \big|_{\mu_W}
  \pm 0.01 \big|_{m_t} \right) \times 10^{-11} \GeV ,
\\
  \im A_2 &
  = \left(-7.09
  \pm 0.23 \big|^\text{ME}_\text{stat}
  \pm 0.43 \big|^\text{ME}_\text{syst}
  \pm 0.37 \big|_\text{CKM}
  \; {}^{+0.34}_{-1.01} \big|_{\mu_c}
  \; {}^{+1.34}_{-1.00} \big|_{\mu_W}
  \pm 0.12 \big|_{m_t} \right) \times 10^{-13} \GeV ,
\end{align}
where NNLO QCD corrections have been included in EWP parts
\cite{Aebischer:2019mtr}. The statistical uncertainties due to the matrix
elements (ME, stat) were determined including the available correlations
for $I=0$, whereas the systematic ones (ME, syst) are based on the overall
$15.7\,\%$ for $I=0$ and $(3-6)\,\%$ for $I=2$, as estimated by RBC-UKQCD
in \cite{Abbott:2020hxn} and \cite{Blum:2015ywa}, respectively.
For comparison, these values are very close to the RBC-UKQCD predictions
$\re A_0 = 24.44 \times 10^{-8} \GeV$,
$\re A_2 =  1.22 \times 10^{-8} \GeV$,
$\im A_0 = -5.70 \times 10^{-11} \GeV$,
$\im A_2 = -6.81 \times 10^{-13} \GeV$,
from Eqs.$\,$(77a, 85, 90) \cite{Abbott:2020hxn} and Eq.$\,$(64)
\cite{Blum:2015ywa}, respectively\footnote{We use here $h=1$ as opposed to
RBC-UKQCD collaboration that uses $h=\sqrt{3/2}$.}. The scale uncertainties
are obtained by varying $\mu_c\in[1.0,\,3.0]\GeV$ and $\mu_W\in[50,\,140]\GeV$
for the NLO expressions, shown in \reffig{fig:mu-dep}. Note that we use
$m_t(\mu_W)$, and hence the $\mu_W$
variation includes the top-mass scheme dependence. We emphasize
that the $\mu_W$ uncertainty for $\im A_{0,2}$, and $\epe$, is very
conservative, because we actually include here partial NNLO QCD corrections
to EWPs \cite{Buras:1999st}, which remove the implicit $\mu_W$ dependence
associated with the top-quark mass and some of the explicit $\mu_W$ dependence
as well, see also \cite{Aebischer:2019mtr} for more details. The parametric
uncertainty due to the input value for the top-quark mass in
\reftab{tab:num-in-WC} is denoted by ``$m_t$''.

\begin{table}
\centering
\renewcommand{\arraystretch}{1.3}
\begin{tabular}{|c|rr|rrrr|rrrr|}
\hline
& $Q_1$ & $Q_2$
& $Q_3$ & $Q_4$ & $Q_5$ & $Q_6$
& $Q_7$ & $Q_8$ & $Q_9$ & $Q_{10}$
\\
\hline\hline
  $\re A_0$
&  12.7 & 95.8 & 0.2 & 2.4 & 1.1 & $-12.0$ & 0.1 & $-0.2$ & 0.0 & 0.0
\\
  $\re A_2$
& $-27.4$ & 128.6 & 0.0 & 0.0 & 0.0 & 0.0 & 0.3 & $-1.5$ & 0.0 & 0.0
\\
\hline
  $\im A_0$
&  0.0 & 0.0 & $-2.7$ & $-16.9$ & $-7.5$ & 121.8 & $-0.2$ & 3.4 & 1.8 & 0.4
\\
  $\im A_2$
& 0.0 & 0.0 & 0.0 & 0.0 & 0.0 & 0.0 & $-5.8$ & 120.2 & $-18.0$ & 3.6
\\
\hline
\end{tabular}
\renewcommand{\arraystretch}{1.0}
\caption{The contribution in \% of each operator to $\re A_{0,2}$ and
  $\im A_{0,2}$ at $\mu_{0,2}$.
}
\label{tab:ampl-composition}
\end{table}

The various relative contributions of the operators to $\re A_{0,2}$ and
$\im A_{0,2}$ are listed in \reftab{tab:ampl-composition} when using
$\mu_{0,2}$. These numbers show that $\re A_{0,2}$ are dominated by
the current-current operators. In $\re A_0$ the $Q_2$ dominates with
almost 96\%, whereas the $Q_1$ and $Q_6$ contributions of about
12\% cancel each other and there are subleading $2\%$ and $1\%$ contributions
from $Q_4$ and $Q_5$. In $\re A_2$ the $Q_2$ of $129\%$ and the $Q_1$ of
$27\%$ enter with opposite signs and there is a subleading contribution
from $Q_8$ of $-1.5$\%. On the other hand the $\im A_0$ is dominated by
QCDP operators, where the $121$\% contribution of $Q_6$ is mainly
reduced  by $Q_4$ and $Q_5$. The $\im A_2$ is dominated by EWPs, in
particular by $122\%$ due to $Q_8$, which is partially cancelled by
$Q_9$. The $5\%$ corrections from $Q_7$ and $Q_{10}$ cancel each other.

In the SM $\epe$ receives contributions from QCDP and EWP via the $I =0$
matrix elements and from EWP via the $I=2$ matrix elements, that exhibit
quite some hierarchies as can be seen in \eqref{eq:me-I=0-mu0} and
\eqref{eq:me-I=2-mu2}, respectively. These hierarchies are strongly
counteracted by those present in the Wilson coefficients $y_i$ at the
two scales $\mu_0 = 4.006\GeV$ and $\mu_2 = 3.0\GeV$, where we evaluate
$\im \wT{A}_0$ and $\im A_2$, respectively. This is illustrated by the
following semi-analytic results of $\epe$ that include the NNLO QCD
corrections to EWPs \cite{Buras:1999st}
\begin{equation}
  \label{eq:semi-num-2}
\begin{aligned}
  \frac{\varepsilon'}{\varepsilon} =
  \im \lambda_t \cdot \Big\{ &
    a (1 - \OmHatEff) \big[
     8.12 \langle Q_3 \rangle_0
  - 23.26 \langle Q_4 \rangle_0
  +  5.47 \langle Q_5 \rangle_0
  - 23.72 \langle Q_6 \rangle_0 \big]
\\ &
  -  0.06 \langle Q_7 \rangle_0
  +  0.25 \langle Q_8 \rangle_0
  -  3.85 \langle Q_9 \rangle_0
  +  0.66 \langle Q_{10} \rangle_0
\\ &
  +  1.42 \langle Q_7 \rangle_2
  -  6.45 \langle Q_8 \rangle_2
  + 70.33 \langle Q_9 \rangle_2 \Big\} .
\end{aligned}
\end{equation}
Here the experimental values of $\re A_{0,2}$ have been used only in the
denominator of \eqref{eq:eprime}. As a remark on the side, we
note that at the scale $\mu_2 = 3\GeV$ the relations $\langle Q_{3,4,5,6}
\rangle_2 = 0$ hold because the RBC-UKQCD calculations so far do not include
isospin-breaking corrections, neither due to quark masses nor quark charges.
In consequence no such contributions appear in~\eqref{eq:semi-num-2}.
Moreover isospin relations \eqref{eq:isospin-2-rel} have been used to
substitute $\langle Q_{1,2,10} \rangle_2 \to \langle Q_9 \rangle_2$.
As mentioned before in \refsec{sec:analytic-formula},
a straight-forward application of the NLO QCD$\,\times\,$QED RG
equations to the matrix elements to evolve them to some different scale
will generate nonvanishing $\langle Q_{3,4,5,6} \rangle_2$, because the
RG flow includes isospin-breaking effects from quark charges. The effect
on $\epe$ is discussed in further detail in \refapp{app:RG-isopin-breaking},
where we provide the analogous result to \eqref{eq:semi-num-2} at
$\mu = 1.3\GeV$.

The hierarchy of the Wilson coefficients signaled for instance
by large coefficients in front of $\langle Q_{3,4} \rangle_0$ is
strongly counteracted by a hierarchy in the hadronic matrix elements
modifying the pattern of the various contributions:
\begin{equation}
  \label{eq:semi-num-3}
\begin{aligned}
  \frac{\varepsilon'}{\varepsilon} =
  \im \lambda_t \cdot \Big\{ &
    a (1 - \OmHatEff) \big[
  - 0.57 \big|_{3,0}
  - 3.51 \big|_{4,0}
  - 1.56 \big|_{5,0}
  +25.33 \big|_{6,0} \big]
\\ &
  - 0.03 \big|_{7,0}
  + 0.70 \big|_{8,0}
  + 0.37 \big|_{9,0}
  + 0.07 \big|_{10,0}
\\ &
  + 0.33 \big|_{7,2}
  - 6.91 \big|_{8,2}
  + 0.83 \big|_{9,2} \Big\}\,,
\end{aligned}
\end{equation}
where the ``$|_{i,I}$'' indicate the origin of the contribution.
This shows much clearer the relevance of $\langle Q_6 \rangle_0 \sim \bsi$
and $\langle Q_8 \rangle_2 \sim \bei$ for $\epe$ and to some extend
$\langle Q_4 \rangle_0$. Eventually
\begin{align}
  \label{eq:semi-num-4}
  \frac{\varepsilon'}{\varepsilon} =
  \im \lambda_t \cdot \Big\{ &
    19.69 \, a (1 - \OmHatEff) \big|_{\text{QCDP},0}
    + 1.11 \big|_{\text{EWP},0}
    - 5.75 \big|_{\text{EWP},2} \Big\}
\end{align}
shows the contributions of QCDP in $I=0$ and the partial cancellation
of EWP contributions from $I=0$ and $I=2$. Note that
this statement is scale dependent, i.e. at some other scales $\mu_{0,2}$
the composition changes slightly due to RG flow.

The final result for $a = 1.017$, using $\OmHatEff^{(8)} = 0.17 \pm 0.9$
in the octet scheme \eqref{eq:OmEff-8}, with NNLO QCD in EWP and other parameters
as collected in \reftab{tab:num-input} and \reftab{tab:num-in-WC} is
\begin{align}
  (\epe)^{(8)} &
  = \left(
  17.4
  \pm 2.3 \big|^\text{ME}_\text{stat}
  \pm 4.9 \big|^\text{ME}_\text{syst}
  \pm 2.6 \big|_{\OmHatEff}
  \pm 1.0 \big|_{\im \lambda_t}
  \; {}^{+0.2}_{-0.6} \big|_{\mu_c}
  \; {}^{+0.4}_{-0.6} \big|_{\mu_W}
  \pm 0.1 \big|_{m_t}
\right) \times 10^{-4}
\notag
\\[1mm] &
  \label{eq:epe-final-8}
  = \left( 17.4 \pm 6.1 \right) \times 10^{-4}.
\end{align}
Considering the new value $\OmHatEff^{(9)} = 0.29 \pm 0.07$ from
the nonet scheme \eqref{eq:OmEff-9} we find:
\begin{align}
  (\epe)^{(9)} &
  = \left(
  13.9
  \pm 2.0 \big|^\text{ME}_\text{stat}
  \pm 4.2 \big|^\text{ME}_\text{syst}
  \pm 2.0 \big|_{\OmHatEff}
  \pm 0.8 \big|_{\im \lambda_t}
  \; {}^{+0.2}_{-0.5} \big|_{\mu_c}
  \; {}^{+0.5}_{-0.7} \big|_{\mu_W}
  \pm 0.1 \big|_{m_t}
\right) \times 10^{-4}
\notag
\\[1mm] &
  \label{eq:epe-final-9}
  = \left( 13.9 \pm 5.2 \right) \times 10^{-4}.
\end{align}

There is a statistical error due to the matrix elements from the lattice, based
on covariance matrices for $I=0$, propagated with Monte Carlo methods as well
as individually available statistical errors for $I=2$ matrix elements. The
systematic uncertainty due to various sources related to the lattice approach
is entirely dominated by the $15.7\,\%$ systematic error of $\langle Q_6
\rangle_0$ in $\im A_0$. The isospin-breaking corrections
to QCDP from ChPT, summarized in $\OmHatEff$ in \refeq{OM+},
contribute a relative uncertainty of $15\,\%$. There is an overall relative
uncertainty of $5.5\,\%$ from $\im \lambda_t$ due to the CKM input.

The NNLO QCD corrections to EWPs lead to a decrease of $\epe$
\cite{Aebischer:2019mtr} and without them the central value would be
$\epe = 18.1 \times 10^{-4}$. Since our numerical input and the treatment
of short-distance contributions differs slightly from RBC-UKQCD our central
value does not agree exactly with their prediction $(\epe)_\text{RBC-UKQCD}
=  21.7\times 10^{-4}$ \cite{Abbott:2020hxn}, such that after setting
$a = 1.0$, $\OmHatEff = 0.0$, and using only NLO QCD EWP we obtain slightly
higher $\epe = 22.6 \times 10^{-4}$, but well within the uncertainties.
The inclusion of NNLO QCD EWP reduces this to $\epe = 21.8 \times 10^{-4}$.

In our prediction we made only use of the experimental values
$\re A_{0,2}|_\text{exp}$ in the denominator of \eqref{eq:eprime}.
As proposed in \cite{Buras:2015yba}, in addition also in the numerator
one of the $I=0$ and one of the $I=2$ matrix elements could be
eliminated in favour of $\re A_{0,2}|_\text{exp}$ to improve the
accuracy in the framework of the SM. Here we did not adapt this strategy,
because in agreement with the RBC-UKQCD collaboration \cite{Abbott:2020hxn},
we did not find evidence for a substantial improvement when employing it
to the $I=0$ amplitude. It must be also noted that this strategy leads to
a slightly reduced value of $\im A_0$ compared to the result without the
additional information from $\re A_0|_\text{exp}$.

The ``$\Delta I = 1/2$ rule'' is given by the ratio
\begin{align}
  \frac{\re A_0}{\re A_2} &
  = 20.0
  {}^{+2.3}_{-2.1} \big|^\text{ME}_\text{stat}
  \pm 3.3 \big|^\text{ME}_\text{syst}
  \; {}^{+0.3}_{-0.2} \big|_{\mu_c}
  \; {}^{+1.4}_{-1.2} \big|_{\mu_W} \,,
\end{align}
and agrees with the experimental result $22.35 \pm 0.05$.
Our value almost coincides with the RBC-UKQCD prediction. The RBC-UKQCD
lattice results show that QCD dynamics, present dominantly in current-current
operators, is responsible for this large ratio  thereby confirming the
findings within DQCD obtained many years ago \cite{Bardeen:1986vz,
  Buras:2014maa}.

%
%
%
\section{BSM master formula}
\label{sec:NP}

In this section we report the updated master formula coefficients describing
the new physics effects beyond the SM (BSM) in $\epe$,
\begin{align}
  \frac{\varepsilon'}{\varepsilon} &
  = \left(\frac{\varepsilon'}{\varepsilon}\right)_\text{SM}
  + \;\; \left(\frac{\varepsilon'}{\varepsilon}\right)_\text{BSM}\,,
\end{align}
which were first presented in \cite{Aebischer:2018quc, Aebischer:2018csl}.
The BSM contribution to $\epe$ is given by the weight factors $P_i$ for each
Wilson coefficient $C_i(\muEW)$ of the operators and their chirality-flipped
counterparts listed in \reftab{tab:P_i}. The $P_i(\muEW)$ contain the
information of the RG evolution from the low-energy scale $\mu$ to the
electroweak (EW) scale $\muEW$ and are linearly dependent on the hadronic
matrix elements of the operators at the scale $\mu$, such that the
$\mu$-dependence cancels. The master formula takes the simple form
\begin{align}
  \left(\frac{\varepsilon'}{\varepsilon}\right)_\text{BSM} &
  = \; \sum_i P_i(\muEW) \im \Big[ C_i(\muEW) - C_i^\prime(\muEW)\Big] \,,
\end{align}
with the $N_f=5$ effective Hamiltonian
\begin{equation}
  \mathcal{H}^{(5)}_{\Delta S =1}
  = -\sum_i \frac{C_i(\muEW)}{(1\,\tev)^2} Q_i\,,
\end{equation}
leading to dimensionless Wilson coefficients and $P_i$ factors.
The sum runs over all Wilson coefficients of the operators in
\reftab{tab:P_i}. These operators are a complete basis for
non-leptonic $\Delta S = 1$ transitions in the absence of any other light
degrees of freedom \cite{Aebischer:2018csl}. The Wilson coefficients and
their weight factors are evaluated at the particular value $\muEW = 160\GeV$
of the EW scale. For more details we refer to \cite{Aebischer:2018quc,
Aebischer:2018csl}.

In \reftab{tab:P_i} we summarize the updated $P_i$ factors after taking
into account the most recent $I=0$ matrix elements reported by RBC-UKQCD
\cite{Abbott:2020hxn}. \reftab{tab:P_i} has been obtained by taking into
account the tree-level matching \cite{Aebischer:2015fzz} and one-loop
running \cite{Aebischer:2017gaw} below the EW scale using the public codes
\texttt{wilson}~\cite{Aebischer:2018bkb} and \texttt{WCxf}
\cite{Aebischer:2017ugx}. Only the $P_i$ factors of operators in Class A
are affected by this change, since they depend exclusively on matrix elements
of the SM operators. In all other classes the $P_i$'s depend on matrix
elements of BSM operators or the chromomagnetic operator $Q_{8g}$ and
remain unchanged. The central values as well as statistical
and systematic uncertainties of the $I=0,2$ matrix elements of all operators
are listed in \reftab{tab:me-values} at the common scale $\mu = 1.3 \GeV$.

\begin{table}
\centering
\renewcommand{\arraystretch}{1.02}
\begin{tabular}{|clrr|}
\hline
  Class & $Q_i$       & $P_i$ & $\frac{\Lambda}{\text{TeV}}$
\\
\hline\hline
  \multirow{18}{*}{A)}
& $Q_{VLL}^u = (\bar s^i \gamma_\mu P_L d^i)(\bar u^j \gamma^\mu P_L u^j)$                & $-3.3  \pm 0.8$  & 57  \\
& $Q_{VLR}^u = (\bar s^i \gamma_\mu P_L d^i)(\bar u^j \gamma^\mu P_R u^j)$                & $-124  \pm 11$   & 351 \\
& $\wT{Q}_{VLL}^u = (\bar s^i \gamma_\mu P_L d^j)(\bar u^j \gamma^\mu P_L u^i)$           & $1.1   \pm 1.2$  & 32  \\
& $\wT{Q}_{VLR}^u = (\bar s^i \gamma_\mu P_L d^j)(\bar u^j \gamma^\mu P_R u^i)$           & $-430  \pm 40$   & 656 \\[0.1cm]
& $Q_{VLL}^d = (\bar s^i \gamma_\mu P_L d^i)(\bar d^j \gamma^\mu P_L d^j)$                & $1.8   \pm 0.5$  & 42  \\
& $Q_{VLR}^d = (\bar s^i \gamma_\mu P_L d^i)(\bar d^j \gamma^\mu P_R d^j)$                & $117   \pm 11$   & 342 \\
& $Q_{SLR}^d = (\bar s^i P_L d^i)(\bar d^j P_R d^j)$                                      & $204   \pm 20$   & 451 \\[0.1cm]
& $Q_{VLL}^s = (\bar s^i \gamma_\mu P_L d^i)(\bar s^j \gamma^\mu P_L s^j)$                & $0.1   \pm 0.1$  & 7   \\
& $Q_{VLR}^s = (\bar s^i \gamma_\mu P_L d^i)(\bar s^j \gamma^\mu P_R s^j)$                & $-0.17 \pm 0.04$ & 12  \\
& $Q_{SLR}^s = (\bar s^i P_L d^i)(\bar s^j P_R s^j)$                                      & $-0.4  \pm 0.1$  & 19  \\[0.1cm]
& $Q_{VLL}^c = (\bar s^i \gamma_\mu P_L d^i)(\bar c^j \gamma^\mu P_L c^j)$                & $0.5   \pm 0.1$  & 22  \\
& $Q_{VLR}^c = (\bar s^i \gamma_\mu P_L d^i)(\bar c^j \gamma^\mu P_R c^j)$                & $0.8   \pm 0.1$  & 28  \\
& $\wT{Q}_{VLL}^c = (\bar s^i \gamma_\mu P_L d^j)(\bar c^j \gamma^\mu P_L c^i)$           & $0.7   \pm 0.1$  & 26  \\
& $\wT{Q}_{VLR}^c = (\bar s^i \gamma_\mu P_L d^j)(\bar c^j \gamma^\mu P_R c^i)$           & $1.3   \pm 0.2$  & 35  \\[0.1cm]
& $Q_{VLL}^b = (\bar s^i \gamma_\mu P_L d^i)(\bar b^j \gamma^\mu P_L b^j)$                & $-0.33 \pm 0.03$ & 18  \\
& $Q_{VLR}^b = (\bar s^i \gamma_\mu P_L d^i)(\bar b^j \gamma^\mu P_R b^j)$                & $-0.22 \pm 0.03$ & 14  \\
& $\wT{Q}_{VLL}^b = (\bar s^i \gamma_\mu P_L d^j)(\bar b^j \gamma^\mu P_L b^i)$           & $0.3   \pm 0.1$  & 17  \\
& $\wT{Q}_{VLR}^b = (\bar s^i \gamma_\mu P_L d^j)(\bar b^j \gamma^\mu P_R b^i)$           & $0.4   \pm 0.1$  & 19  \\
\hline
  \multirow{11}{*}{B)}
& $Q_{8g} \;\;\, = m_s (\bar s \sigma^{\mu\nu} T^a P_L d)\, G^{a}_{\mu\nu}$               & $-0.35 \pm 0.12$ & 18 \\ [0.1cm]
& $Q_{SLL}^s = (\bar s^i P_L d^i)(\bar s^j P_L s^j)$                                      & $0.05  \pm 0.02$ & 7  \\
& $Q_{TLL}^s = (\bar s^i \sigma_{\mu\nu} P_L d^i)(\bar s^j \sigma^{\mu\nu} P_L s^j)$      & $-0.14 \pm 0.05$ & 12 \\ [0.1cm]
& $Q_{SLL}^c = (\bar s^i P_L d^i)(\bar c^j P_L c^j)$                                      & $-0.26 \pm 0.09$ & 16 \\
& $Q_{TLL}^c = (\bar s^i \sigma_{\mu\nu} P_L d^i)(\bar c^j \sigma^{\mu\nu} P_L c^j)$      & $-0.15 \pm 0.05$ & 12 \\
& $\wT{Q}_{SLL}^c = (\bar s^i P_L d^j)(\bar c^j P_L c^i)$                                 & $-0.23 \pm 0.07$ & 15 \\
& $\wT{Q}_{TLL}^c = (\bar s^i \sigma_{\mu\nu} P_L d^j)(\bar c^j \sigma^{\mu\nu} P_L c^i)$ & $-5.9  \pm 1.9$  & 76 \\ [0.1cm]
& $Q_{SLL}^b = (\bar s^i P_L d^i)(\bar b^j P_L b^j)$                                      & $-0.35 \pm 0.12$ & 18 \\
& $Q_{TLL}^b = (\bar s^i \sigma_{\mu\nu} P_L d^i)(\bar b^j \sigma^{\mu\nu} P_L b^j)$      & $-0.11 \pm 0.03$ & 10 \\
& $\wT{Q}_{SLL}^b = (\bar s^i P_L d^j)(\bar b^j P_L b^i)$                                 & $-0.34 \pm 0.11$ & 18 \\
& $\wT{Q}_{TLL}^b = (\bar s^i \sigma_{\mu\nu} P_L d^j)(\bar b^j \sigma^{\mu\nu} P_L b^i)$ & $-13.4 \pm 4.5$  & 115 \\
\hline
  \multirow{4}{*}{C)}
& $Q_{SLL}^u = (\bar s^i P_L d^i)(\bar u^j P_L u^j)$                                      & $74    \pm 16$   & 272 \\
& $Q_{TLL}^u = (\bar s^i \sigma_{\mu\nu} P_L d^i)(\bar u^j \sigma^{\mu\nu} P_L u^j)$      & $-162  \pm 36$   & 402 \\
& $\wT{Q}_{SLL}^u = (\bar s^i P_L d^j)(\bar u^j P_L u^i)$                                 & $-15.6 \pm 3.3$  & 124 \\
& $\wT{Q}_{TLL}^u = (\bar s^i \sigma_{\mu\nu} P_L d^j)(\bar u^j \sigma^{\mu\nu} P_L u^i)$ & $-509  \pm 108$  & 713 \\
\hline
  \multirow{2}{*}{D)}
& $Q_{SLL}^d = (\bar s^i P_L d^i)(\bar d^j P_L d^j)$                                      & $-87   \pm 16$   & 295 \\
& $Q_{TLL}^d = (\bar s^i \sigma_{\mu\nu} P_L d^i)(\bar d^j \sigma^{\mu\nu} P_L d^j)$      & $191   \pm 35$   & 436 \\
\hline
  \multirow{2}{*}{E)}
& $Q_{SLR}^u = (\bar s^i P_L d^i)(\bar u^j P_R u^j)$                                      & $-266   \pm 21$  & 515 \\
& $\wT{Q}_{SLR}^u = (\bar s^i P_L d^j)(\bar u^j P_R u^i)$                                 &  $-60   \pm 5$   & 244 \\
\hline
\end{tabular}
\renewcommand{\arraystretch}{1.0}
\caption{
  Updated $P_i$ coefficients entering the master formula for NP effects in $\epe$.
}
  \label{tab:P_i}
\end{table}

The changes are moderate of not more than 30\% for operators that contribute
directly to $K\to\pi\pi$, whereas changes can be larger for those operators
(with $s,c,b$-quarks) that enter via RG running from the EW scale down to
the low-energy scale and have smaller coefficients. The last column of
\reftab{tab:P_i} shows the suppression scale $\Lambda$ that would generate
$(\epe)_\text{BSM} = 10^{−3}$ for $C_i = 1/\Lambda^2$, assuming the presence
of only this particular operator. For comparison, the theory uncertainty of
the SM prediction \eqref{eq:epe-final-8} is about $0.6 \times 10^{-3}$. The scale
$\Lambda$ is strongly dependent on the uncertainties of the matrix elements,
which did not all decrease in the latest RBC-UKQCD predictions. A comparison
to the previous values \cite{Aebischer:2018quc} shows a slight increase of
$\Lambda$ for the first seven operators, which contribute directly to
$K\to\pi\pi$. In general $\Lambda$ also increases for the remaining Class A
operators, with a few exceptions, pushing the NP scale also in these cases up,
even though they are entering only via RG mixing. This shows that the new
results for the matrix elements from RBC-UKQCD will lead to stronger bounds
on CP violation beyond the SM.

Eventually we point out that the large increase of the central value of
$(\epe)_\text{SM}$ in the SM from $\sim (1 - 2) \times 10^{-4}$ with the
2015 RBC-UKQCD results to $\sim 14 \times 10^{-4}$ with the 2020 results
constitutes one order of magnitude and hence has significant
impact on excluded regions of parameter spaces of BSM scenarios. The 2015
SM predictions \cite{Bai:2015nea, Buras:2015yba, Kitahara:2016nld}
suggested a strong anomaly with a constructive $(\epe)_\text{BSM} \approx
(5-15) \times 10^{-4}$ to reach agreement with the experimental value
$(\epe)_\text{exp} = (16.6 \pm 2.3) \times 10^{-4}$. Contrary, the
$(\epe)_\text{SM}$ predictions based on 2020 results do not show anymore
an anomaly, but allow now for both, a constructive and destructive interference,
that can be still sizable in view of the large theory uncertainties
\begin{align}
   -4 \times 10^{-4}
   \;\lesssim\; \left( \frac{\varepsilon'}{\varepsilon} \right)_\text{BSM}
   \;\lesssim\; + 10 \times 10^{-4}
\end{align}
as a rough $1\,\sigma$ range. The complete error propagation can be
obtained properly for general BSM scenarios with the master formula, which
is implemented in the public code \texttt{flavio}~\cite{flavio,
Straub:2018kue}. Despite the large uncertainties, $\epe$ was and remains one of
the strongest constraints
on CP violation in the quark-flavour sector, as has been shown for different
BSM scenarios in the past. The BSM studies based on the 2015 SM predictions
of $\epe$ used mostly the working hypothesis of a constructive
$(\epe)_\text{BSM}$ of similar size, see references in \cite{Aebischer:2019mtr},
and the obtained conclusions for $0 < (\epe)_\text{BSM}$ are still mostly valid.

%
%
%
\section{Summary and outlook}
\label{sec:4}

Our final result for $\epe$ differs significantly from
the one of the RBC-UKQCD collaboration but in view of large uncertainties
in both results they are in agreement with each other and with experiment.
Our result is in good agreement with the one of ChPT in \eqref{Pich}
but it should be clarified whether it is a pure numerical coincidence
or indeed QCD dynamics, that enhancing the parameter $\bsi$ over unity
in LQCD and in ChPT is the same.

The recent advances in LQCD allow us to hope that in the coming years we should
be able to have a value of $\epe$ within the SM with a comparable error to the
experimental one. In order to reach this goal and thereby to obtain an assessment
on the allowed room for NP contributions to $\epe$ it is important to perform a
number of steps:
\begin{itemize}
\item A more precise determination of $\langle Q_6(\mu_0)\rangle_0$ or
  $\bsi(\mu_0)$.
  At least a second LQCD collaboration should calculate $\epe$, in order to
  confirm the large enhancement of $\bsi$ found by RBC-UKQCD that has not been
  identified in DQCD. Also the errors in other  matrix elements should be decreased.
\item A more precise determination of $\OmHatEff$. In particular in LQCD
  calculations isospin-breaking corrections and $\ord(\alE)$ corrections
  in hadronic matrix elements required for the removal of renormalization scheme
  dependence at this order should be taken into account.
  The present status is summarized in \cite{Giusti:2018guw}.
\item A more precise determination of the short distance contributions,
  especially in the QCD penguin sector, which in the context of the
  RBC-UKQCD analysis will decrease the sensitivity to the matching scale
  $\mu_c$. Despite the fact that the NNLO analysis of QCD
  corrections to EWP contributions practically removed the sensitivity of
  $\epe$ to the renormalization scheme of the top-quark mass and $\mu_W$,
  our analysis shows that the significant
  $\mu_c$ uncertainty in the EWP sector still has to be removed through the
  matching of $N_f=4$ to $N_f=3$ effective theory at the NNLO level.
\item The computation of the BSM $K\to\pi\pi$ hadronic matrix elements of
  four-quark operators by lattice QCD, which are presently known only from
  the DQCD approach \cite{Aebischer:2018rrz}.
\end{itemize}

Several BSM analyses of $\epe$ have been performed, which are collected in
\cite{Aebischer:2019mtr}. A recent example of a $Z'$ model with explicit
gauge anomaly cancellation has been discussed in \cite{Aebischer:2019blw}.
Furthermore leptoquark models, except the $U_1$
model, would  not be able to explain large deviations of the SM value from the
data due to  constraints coming from rare $K$ decays \cite{Bobeth:2017ecx}.
This underlines the importance of correlations of $\epe$ with other observables
in NP scenarios. The new SM value in~\eqref{BG20} removes the difficulties
of leptoquark models pointed out in \cite{Bobeth:2017ecx}, but these problems
could return with an improved analyses of $\epe$ within the SM.

Furthermore the lessons from the SMEFT analysis in \cite{Aebischer:2018csl}
should be useful in this respect. Such general analyses allow to take into
account constraints from other processes such as collider processes,
electroweak precision tests, neutral meson mixing as well as electric dipole
moments. Finally the master formula for $\epe$ presented in
\cite{Aebischer:2018quc} valid for any BSM scenario should facilitate
the derivation of constraints on CP-violating phases beyond
the SM imposed by $\epe$. In this respect we point out that also $\re A_2$
has a very precise SM prediction and can
be predicted rather precisely also in BSM scenarios, providing thus
a second observable besides $\epe$ to constrain also real parts of
the Wilson coefficients of non-leptonic $\Delta S = 1$ operators.

%
%
%
\section*{Acknowledgements}

We thank Jean-Marc G{\'e}rard for discussions and comments on the manuscript.
We also thank Christopher Kelly and Chris Sachrajda for informative
email exchanges related to the RBC-UKQCD result and Maria Cerd{\`a}-Sevilla
for discussions on NNLO QCD corrections to QCDP.
J.A. acknow\-ledges financial support from the Swiss National Science
Foundation (Project No. P400P2\_183838).
The research of A.J.B was supported by the Excellence Cluster ORIGINS,
funded by the Deutsche Forschungsgemeinschaft (DFG, German Research Foundation)
under Germany's Excellence Strategy – EXC-2094 – 390783311.

%
%
%
\appendix

%
%
%
\section{Hadronic matrix elements}
\label{app:hadr-me}

Here we collect the input for the $K\to (\pi\pi)_I$ isospin $I=0,2$ hadronic
matrix elements
\begin{align}
  \langle Q_i \rangle_I &
  \equiv \bra{(\pi\pi)_I} Q_i \ket{K}
\end{align}
of the relevant operators in the traditional SM basis \cite{Buras:1993dy}.
They are given for the \msbar{} scheme in order to combine them with the
Wilson coefficients in \refapp{app:wilson-coeffs} at the scale chosen
by RBC-UKQCD. In addition we provide these matrix elements together with
the complete set of non-leptonic $\Delta S = 1$ operators beyond the SM in
\reftab{tab:me-values} at the common scale $\mu = 1.3$~GeV. These results
can be used for new physics studies using the master formula of $\epe$ in
\cite{Aebischer:2018quc, Aebischer:2018csl} that we updated in
\refsec{sec:NP}.

\begin{table}
\centering
\renewcommand{\arraystretch}{1.3}
\begin{tabular}{
  |@{$\;\;$}c |
  *{1}{D{.}{.}{12}} *{1}{D{.}{.}{12}} |
  *{1}{D{.}{.}{12}} *{1}{D{.}{.}{12}} |}
\hline
  $Q_i$
& \multicolumn{2}{c|}{$\langle Q_i \rangle_0$}
& \multicolumn{2}{c|}{$\langle Q_i \rangle_2$}
\\
\hline\hline
  \multicolumn{5}{|c|}{SM operators}
\\
\hline
& \multicolumn{1}{c}{QCD$\,\times\,$QED} & \multicolumn{1}{c|}{QCD}
& \multicolumn{1}{c}{QCD$\,\times\,$QED} & \multicolumn{1}{c|}{QCD}
\\
\hline
  $Q_1$     & -0.065(17)(10)   &  -0.065(17)(10) & 0.0087(2)(5)    & 0.0085(2)(5)
\\
  $Q_2$     &  0.087(13)(14)   &   0.087(13)(14) & 0.0085(2)(5)    & 0.0085(2)(5)
\\
\hline
  $Q_3$     & -0.075(57)(12)   & -0.075(57)(12)  & 0.0000          & 0
\\
  $Q_4$     &  0.093(51)(15)   &  0.093(52)(15)  & -0.0003         & 0
\\
  $Q_5$     & -0.120(53)(19)   & -0.121(53)(19)  & 0.0002          & 0
\\
  $Q_6$     & -0.641(46)(101)  & -0.644(46)(101) & 0.0011          & 0
\\
\hline
  $Q_7$     &  0.217(16)(34)   &  0.216(16)(34)  & 0.0996(68)(30)  & 0.0989(68)(30)
\\
  $Q_8$     &  1.583(30)(249)  &  1.581(30)(249) & 0.684(19)(41)   & 0.683(19)(41)
\\
  $Q_9$     & -0.059(17)(9)    & -0.061(17)(9)   & 0.0132(3)(8)    & 0.0128(3)(8)
\\
  $Q_{10}$  &  0.092(18)(14)   &  0.092(18)(14)  & 0.0130(3)(8)    & 0.0128(3)(8)
\\
\hline
  $Q_{8g}$  & -0.013(4)        &                 & 0               & 0
\\
\hline\hline
  \multicolumn{5}{|c|}{Beyond the SM operators}
\\
\hline
  $Q_1^{\text{SLL},u}$  & -0.005(1)  & & -0.0030(6) &
\\
  $Q_2^{\text{SLL},u}$  & -0.044(9)  & & -0.031(6)  &
\\
  $Q_3^{\text{SLL},u}$  & -0.371(74) & & -0.262(52) &
\\
  $Q_4^{\text{SLL},u}$  & -0.214(43) & & -0.151(30) &
\\
  $Q_1^{\text{SLL},d}$  & 0.0070(14) & & -0.002(4)  &
\\
  $Q_2^{\text{SLL},d}$  & -0.088(18) & & 0.031(6)   &
\\
  $Q_1^{\text{SLR},u}$  & -0.015(3)  & & 0.0030(6)  &
\\
  $Q_2^{\text{SLR},u}$  & -0.141(28) & & 0.050(10)  &
\\
\hline
\end{tabular}
\renewcommand{\arraystretch}{1.0}
\caption{Numerical values of $K\to\pi\pi$ hadronic matrix elements from the
  literature in units of $\geV^3$ in the \msbar{} scheme at the scale
  $\mu = 1.3\GeV$, with statistical and systematic uncertainties.
  For the SM four-quark operators we provide the
  values obtained by RG equations with NLO QCD$\,\times\,$QED and only NLO QCD.
  The normalization convention is chosen to be $h=1$ for all operators.
}
\label{tab:me-values}
\end{table}

The new results for $I=0$ matrix elements of the SM operators from the
year 2020 are from the RBC-UKQCD lattice collaboration \cite{Abbott:2020hxn}.
They are given at the scale $\mu_0 = 4.006\GeV$ in the $N_f = 2 + 1$ flavour
theory. As the current lattice calculation works in the isospin limit, out
of the ten $\langle Q_{1\ldots10} \rangle_0$ there are only seven
linearly independent (for $h=1$):
\begin{equation}
  \label{eq:me-I=0-mu0}
\begin{aligned}
  \langle Q_1 \rangle_0 & = -0.087(18)(14) , &
  \langle Q_2 \rangle_0 & = +0.120(12)(19) , &
\\
  \langle Q_3 \rangle_0 & = -0.070(50)(11) , &
  \langle Q_5 \rangle_0 & = -0.284(51)(45) , &
  \langle Q_6 \rangle_0 & = -1.068(73)(168) ,
\\
  \langle Q_7 \rangle_0 & = +0.628(19)(99) , &
  \langle Q_8 \rangle_0 & = +2.767(52)(434) . &
\end{aligned}
\end{equation}
The first and second errors are of statistical and systematic origin,
respectively. In particular the statistical error comprises also a
covariance matrix provided in \cite{Abbott:2020hxn} that we use for
the uncertainty propagation in \reftab{tab:me-values} and predictions
of $\epe$. The systematic uncertainty due to various sources in the
lattice approach was estimated to be $15.7\,\%$ (see table XXV
\cite{Abbott:2020hxn}) for each matrix element without providing
correlations. For comparison we show in \reftab{tab:me-values} the previous
results \cite{Bai:2015nea} from the year 2015 as well, which have been
evolved from $\mu = 1.53\GeV$ to $\mu = 1.3\GeV$ for that purpose.

The $I = 2$ matrix elements of the SM operators are also from RBC-UKQCD
\cite{Blum:2015ywa} from the year 2015 for $N_f = 2 + 1$. In particular
we use the results from the RI-SMOM $(\slashed q, \slashed q)$ scheme
(Table XVI) and convert them to the \msbar{} scheme with the scheme
conversion factor Eq.$\,(66)$ in \cite{Blum:2012uk}. The matrix elements
are given at $\mu_2 = 3\GeV$ where they fulfill the isospin relations
\begin{align}
  \label{eq:isospin-2-rel}
    \frac{3}{2} \langle Q_1 \rangle_2 &
  = \frac{3}{2} \langle Q_2 \rangle_2
  = \langle Q_9 \rangle_2
  = \langle Q_{10} \rangle_2,
&
  \langle Q_{3,4,5,6} \rangle_2 & = 0 ,
\end{align}
and reduce to three independent ones (for $h=1$)
\begin{align}
  \label{eq:me-I=2-mu2}
  \langle Q_7 \rangle_2 & = 0.2340(52)(70) , &
  \langle Q_8 \rangle_2 & = 1.072(28)(64) , &
  \langle Q_9 \rangle_2 & = 0.0118(3)(7) ,
\end{align}
where we have increased the systematic uncertainty of the results
of the RI-SMOM $(\slashed q, \slashed q)$ by adding in quadrature
the difference of the results in the RI-SMOM $(\slashed q, \slashed q)$
and the RI-SMOM~$(\gamma, \gamma)$ schemes as given in \cite{Blum:2015ywa}
to account for this additional source of systematic uncertainty.
The RG-evolved results at $\mu = 1.3\GeV$ are given in \reftab{tab:me-values},
see also \cite{Aebischer:2018csl}.

The matrix elements of operators beyond the SM were calculated using DQCD
in \cite{Aebischer:2018rrz}. The single error is of parametric and
systematic origin. The matrix element of the chromo-magnetic dipole operator
$O_{8g}$ has been calculated in \cite{Buras:2018evv, Constantinou:2017sgv}
in 2017/18. Note that we use here the normalization of \cite{Aebischer:2018csl,
Aebischer:2018quc}.

%
%
%
\section{Wilson coefficients}
\label{app:wilson-coeffs}

Here we summarize the $\Delta S = 1$ Wilson coefficients at the various scales
used in our analysis in the NDR-\msbar{} scheme using the NLO RG evolution from
\cite{Buras:1993dy}. The numerical input entering the Wilson coefficients is
fixed to values in \reftab{tab:num-in-WC}. The central values for the threshold
scales at which the top-, bottom and charm quark are subsequently decoupled are
chosen as $\mu_W = m_W$ for $N_f = 6 \to 5$, $\mu_b = 4.2\GeV$ for $N_f = 5
\to 4$ and $\mu_c = 1.3 \GeV$ for $N_f = 4 \to 3$. We employ three-loop running
of $\alS$ including threshold quark mass effects such that $\alS(\mu_c) = 0.3767$
and $1/\alE(\mu_c) = 133.84$ in $N_f = 3$. For simplicity we use in
the threshold corrections for $N_f = 5 \to N_f = 4$ for the bottom-quark
mass the value $m_b = 4.2\GeV$, which agrees very well with latest determinations
of the \msbar{} result $\oL{m}_b(\oL{m}_b) = 4.198 \GeV$ \cite{Aoki:2019cca}.
For the charm-quark mass in the threshold corrections we use $m_c = 1.3\GeV$,
which is close to the \msbar{} result $\oL{m}_c(\oL{m}_c) = 1.27 \GeV$, when
using $\oL{m}_c(3\GeV) = 0.988 \GeV$ \cite{Aoki:2019cca}. We remind that the
threshold corrections enter here for the first time at NLO, hence to be able
to cancel some of the renormalization scheme dependences of the bottom- and
charm-quark masses, one has to go to the NNLO order, as for example done
in \cite{Cerda-Sevilla:2016yzo} in the case of QCD penguins.

\begin{table}
\centering
\renewcommand{\arraystretch}{1.3}
\begin{tabular}{|lll|lll|}
\hline
  Parameter
& Value
& Ref.
&  Parameter
& Value
& Ref.
\\
\hline\hline
  $\alS^{(5)}(m_Z)$           & $0.1181(11)$       & \cite{Tanabashi:2018oca}
& $m_Z$                       & $91.1876$ GeV      & \cite{Tanabashi:2018oca}
\\
  $\alE^{(5)}(m_Z)$           & $1/127.955(10)$    & \cite{Tanabashi:2018oca}
& $m_W$                       & $80.385$ GeV       & \cite{Tanabashi:2018oca}
\\
  $s_W^2 = \sin^2(\theta_W)$  & $0.23126$          & \cite{Tanabashi:2018oca}
& $m_t^\text{pole}$           & $173.1(6)(5)$ GeV  & \cite{Tanabashi:2018oca}
\\
\hline
\end{tabular}
\caption{\small
   Numerical input for Wilson coefficients.
}
  \label{tab:num-in-WC}
\end{table}

The top quark mass $m_t(\mu_t)$ is in the \msbar{} scheme for $\mu_t = \mu_W$,
obtained from the pole mass\footnote{We have interpreted the precisely measured
so-called Monte-Carlo mass as the pole mass, and will include an additional
uncertainty of $\delta m_t = 0.5\GeV$ in \reftab{tab:num-in-WC}, which we
add linearly -- see recent review \cite{Hoang:2020iah} for further details.}
value given in \reftab{tab:num-in-WC}: $m_t(m_t) = 163.5\GeV$
and $m_t(\mu_W) = 173.2 \GeV$. We follow \cite{Aebischer:2019mtr} and include also
important NNLO matching corrections \cite{Buras:1999st} that resolve the NLO
renormalization scheme ambiguities for our choice $\mu_t = \mu_W$ via the
modifications of $y_{7, \ldots, 10}(\mu)$ at the low-energy scale of about
$1.07$, $1.07$, $0.89$ and $0.76$ leading to the NNLO' values in
\reftab{tab:wc-values}, which we adapt in the numerics. For further details
we refer to \cite{Aebischer:2019mtr}. The prime in this indicates that still
small $\ord(\alpha_W\alpha_s\sin^2\theta_W)$ corrections are not included.

\newcommand{\la}{\leftarrow}

\begin{table}
\centering
\renewcommand{\arraystretch}{1.3}
\begin{tabular}{|c|rr|rr|rr|}
\hline
& \multicolumn{2}{c|}{$\mu = 1.3\GeV$}
& \multicolumn{2}{c|}{$\mu = 3.0\GeV$}
& \multicolumn{2}{c|}{$\mu = 4.006\GeV$}
\\
& NLO           & NNLO'
& NLO           & NNLO'
& NLO           & NNLO'
\\
\hline\hline
 $z_1$                & $-0.3938$ & $\la$     & $-0.2368$ & $\la$     & $-0.1984$ & $\la$ \\
 $z_2$                & $ 1.2020$ & $\la$     & $ 1.1096$ & $\la$     & $ 1.0892$ & $\la$ \\
 $z_3 \times 10^2$    & $ 0.4231$ & $\la$     & $-0.3540$ & $\la$     & $-0.4679$ & $\la$ \\
 $z_4 \times 10^2$    & $-1.2693$ & $\la$     & $ 1.5289$ & $\la$     & $ 2.1423$ & $\la$ \\
 $z_5 \times 10^2$    & $ 0.4231$ & $\la$     & $-0.3142$ & $\la$     & $-0.5236$ & $\la$ \\
 $z_6 \times 10^2$    & $-1.2693$ & $\la$     & $ 1.0955$ & $\la$     & $ 1.5460$ & $\la$ \\
 $z_7 \times 10^4$    & $ 0.4780$ & $\la$     & $ 0.8969$ & $\la$     & $ 1.2560$ & $\la$ \\
 $z_8 \times 10^4$    & $ 0     $ & 0         & $-0.9518$ & $\la$     & $-1.0783$ & $\la$ \\
 $z_9 \times 10^4$    & $ 0.4780$ & $\la$     & $ 0.2914$ & $\la$     & $ 0.5490$ & $\la$ \\
 $z_{10} \times 10^4$ & $ 0     $ & 0         & $ 0.7362$ & $\la$     & $ 0.8552$ & $\la$ \\
\hline\hline
 $y_3 \times 10^2$    & $ 2.6958$ & $\la$     & $ 2.0441$ & $\la$     & $ 1.8743$ & $\la$ \\
 $y_4 \times 10^2$    & $-5.4542$ & $\la$     & $-5.3848$ & $\la$     & $-5.3689$ & $\la$ \\
 $y_5 \times 10^2$    & $ 0.5579$ & $\la$     & $ 1.1474$ & $\la$     & $ 1.2634$ & $\la$ \\
 $y_6 \times 10^2$    & $-8.2572$ & $\la$     & $-5.9125$ & $\la$     & $-5.4750$ & $\la$ \\
 $y_7 \times 10^2$    & $-0.0180$ & $-0.0192$ & $-0.0137$ & $-0.0146$ & $-0.0119$ & $-0.0128$ \\
 $y_8 \times 10^2$    & $ 0.0981$ & $ 0.1050$ & $ 0.0622$ & $ 0.0666$ & $ 0.0547$ & $ 0.0585$ \\
 $y_9 \times 10^2$    & $-1.1167$ & $-0.9939$ & $-1.0184$ & $-0.9063$ & $-0.9975$ & $-0.8878$ \\
 $y_{10} \times 10^2$ & $ 0.3981$ & $ 0.3025$ & $ 0.2366$ & $ 0.1798$ & $ 0.1997$ & $ 0.1518$ \\
\hline
\end{tabular}
\caption{\small
  The $\Delta S = 1$ Wilson coefficients at various scales $\mu$ in the
  NDR-\msbar{} scheme for the renormalization scales $\mu_W = \mu_t = m_W$,
  $\mu_b = 4.2\GeV$ and $\mu_c = 1.3\GeV$ using NLO and partial NNLO
  matching results for $y_{7,\ldots,10}$. Further $y_{1,2} = 0$.
}
  \label{tab:wc-values}
\end{table}

%
%
%
\section{Isospin-breaking RG effects}
\label{app:RG-isopin-breaking}

In this appendix we comment on isospin-breaking effects in the RG flow
present in the anomalous dimension matrices that govern the scale dependence
of Wilson coefficients and matrix elements of the operators. They are
due to quark charges present in the definitions of the EWP operators and due
to QED corrections, and are of purely perturbative origin known up to
NLO in QCD$\,\times\,$QED \cite{Buras:1992tc, Buras:1992zv, Ciuchini:1993vr}.
To compute the Wilson coefficients at the low-energy scale in the
$N_f = 3$ theory, the isospin-breaking effects in the RG flow are combined with
initial Wilson coefficients, which contain further isospin-breaking effects of
the SM at the electroweak scale. In the predictions of the amplitudes
$A_{0,2}$ as well as for $\epe$ the low-energy Wilson coefficients
are multiplied by the corresponding matrix elements, which in
principle leads to the cancellation of the renormalization scheme dependence.

The matrix elements contain the dynamics of QCD$\,\times\,$QED at scales
below the low-energy scale, i.e. in the nonperturbative regime of QCD.
In the calculation with lattice methods so far no isospin-breaking corrections
have been included, i.e. a purely isospin-symmetric QCD setup is used by
RBC-UKQCD that neglects effects due to different quark masses $m_u \neq m_d$
as well as QED effects\footnote{In the presence of QED a further complication
consists in a proper subtraction of the infrared divergent virtual corrections
and their combination with the treatment of real photon radiation, see e.g.
\cite{Cirigliano:2003gt}.} due to different quark charges. For this reason,
firstly, scheme dependences between Wilson coefficients and matrix elements
can only cancel for the isosymmetric parts. Secondly, the initial values of
the $I=2$ matrix elements fulfill the isospin relations \eqref{eq:isospin-2-rel}
at the scale $\mu_2 = 3\GeV$, where they have been calculated. These isospin
relations become broken by the full QCD$\,\times\,$QED RG flow when evolving
the matrix elements to a different scale, as can be seen for the example in
\reftab{tab:me-values} at $\mu = 1.3\GeV$, as for example
$3/2 \langle Q_1 \rangle_2 \neq 3/2 \langle Q_2 \rangle_2 \neq
\langle Q_9 \rangle_2 \neq \langle Q_{10} \rangle_2$.

In particular nonvanishing $I=2$ matrix elements $\langle Q_{3,4,5,6} \rangle_2
\neq 0$ are generated at $\mu = 1.3\GeV$ that enter $\epe$ with large Wilson
coefficients. At the scale $\mu = 1.3\GeV$ the analogue result to
\eqref{eq:semi-num-2} becomes
\begin{equation}
  \label{eq:semi-num-1.3}
\begin{aligned}
  \frac{\varepsilon'}{\varepsilon} =
  \im \lambda_t \cdot \Big\{ &
    a (1 - \OmHatEff) \big[
    11.68 \langle Q_3 \rangle_0
  - 23.63 \langle Q_4 \rangle_0
  +  2.42 \langle Q_5 \rangle_0
  - 35.77 \langle Q_6 \rangle_0 \big]
\\ &
  -  0.08 \langle Q_7 \rangle_0
  +  0.45 \langle Q_8 \rangle_0
  -  4.31 \langle Q_9 \rangle_0
  +  1.31 \langle Q_{10} \rangle_0
\\ &
  - 260.96 \langle Q_3 \rangle_2
  + 527.98 \langle Q_4 \rangle_2
  -  54.00 \langle Q_5 \rangle_2
  + 799.32 \langle Q_6 \rangle_2
\\ &
  +  1.86 \langle Q_7 \rangle_2
  - 10.16 \langle Q_8 \rangle_2
  + 96.21 \langle Q_9 \rangle_2
  - 29.28 \langle Q_{10} \rangle_2 \Big\} .
\end{aligned}
\end{equation}
The central values of the matrix elements in \reftab{tab:me-values} lead then
to $(\epe)^{(8)} = 18.1\times 10^{-4}$ and $(\epe)^{(9)} = 14.6\times 10^{-4}$,
respectively. These results differ by about $+0.7\times 10^{-4}$ from the
predictions \eqref{eq:epe-final-8} and \eqref{eq:epe-final-9}, which are rather
small variations in view of larger uncertainties in the prediction of $\epe$.
They might be viewed as an estimate of isospin-breaking corrections due to
quark charges. However, the lacking contributions in the nonperturbative matrix
elements could cancel them in large parts, as one would expect on the basis of
renormalization scheme cancellations. We note that they are unrelated to
isopin-breaking quark-mass effects, which are entirely of nonperturbative origin.

In the case of our preferred choice of calculating $\epe$ at the scales
$\mu_{0,2}$ where RBC-UKQCD has calculated $I=0,2$ matrix elements in the
NDR-\msbar{} scheme, these isospin-breaking corrections are entirely
contained in the Wilson coefficients. At a different scale $\mu$ one might
impose the isospin relations for matrix elements by hand,
but a more consistent possibility is the use of the isospin-conserving parts
of the RG equation only, i.e. the NLO QCD evolution obtained by setting
$\alE = 0$. Then the $I=2$ matrix elements fulfill the isospin relations
exactly, whereas for $I=0$ matrix elements they are slightly broken, because
we have kept the small QCD mixing of EWP into QCDP operators. With this
approach we obtain at $\mu = 1.3\GeV$ the predictions $(\epe)^{(8)} = 17.1
\times 10^{-4}$ and $(\epe)^{(9)} = 13.6\times 10^{-4}$, respectively, differing
by about $-0.3\times 10^{-4}$ from the predictions \eqref{eq:epe-final-8} and
\eqref{eq:epe-final-9} at $\mu_{0,2}$. In particular we have used the
NLO QCD evolution of matrix elements for the semi-analytic equation
\eqref{eq:semi-num-1} and coefficients in \reftab{tab:semi-analytic}.

In conclusion these studies show that nonvanishing $I=2$ matrix elements
of QCDP operators due to isospin-breaking effects can have a non-negligible
impact on $\epe$. The size of such effects is completely unknown at present,
contrary to isospin-breaking effects from $I=0$ matrix elements of QCDP
operators, contained in $\OmHatEff$. Here, the obtained variations are based
on the residual scheme dependence from the isopin-breaking contributions
due to quark charges in perturbation theory, contained in the anomalous
dimensions. Their impact on $\epe$ is numerically subleading compared to
other, currently much larger uncertainties.

%
%
%
\renewcommand{\refname}{R\lowercase{eferences}}

\addcontentsline{toc}{section}{References}

\footnotesize

\bibliographystyle{JHEP}
\bibliography{Bookallrefs}

\end{document}